\documentstyle[prl,aps,twocolumn]{revtex}

\begin{document}

\title{Anomalous Kondo-effect in a Quantum Dot at non-zero bias}
\draft
\author{F. Simmel$^1$, R. H. Blick$^1$, J. P. Kotthaus$^1$,
W. Wegscheider$^2$, and M. Bichler$^2$}
\address{$^1$Center for NanoScience and Sektion Physik,
Ludwig-Maximilians-Universit\"at, Geschwister-Scholl-Platz 1,
80539 M\"unchen, Germany\\
$^2$Walter-Schottky-Institut der Technischen Universit\"at M\"unchen,\\ 
Am Coulombwall, 85748 M\"unchen, Germany}

\date{\today} 
\maketitle

\begin{abstract}
We present measurements on the Kondo-effect in a small quantum dot
connected strongly to one lead and weakly to the other. The conductance of the
dot reveals an offset of the Kondo resonance at zero magnetic field. While
the resonance persists in the negative bias regime, it is suppressed in the
opposite direction. This demonstrates the pinning of the
Kondo resonance to the Fermi levels of the leads.

\end{abstract}

\vspace*{0.8cm}

The Kondo effect is a well-known phenomenon in solid state 
physics~\cite{kondo} -- the hybridization of
conduction electrons with the localized electron spin of a magnetic impurity atom 
in a metal leads to an enhancement of resistivity at low 
temperatures. The current interest stems from the fact that 
mesoscopic devices such as quantum dots allow to study such complex 
solid state phenomena on highly controllable systems~\cite{revdots2}.
The Kondo effect in quantum dots was proposed in early theoretical 
work~\cite{theo1,theo2} and then demonstrated in beautiful experiments 
by Goldhaber-Gordon~{\it et al.}~\cite{davidgg,davidgg2} 
and Cronenwett~{\it et al.}~\cite{crone}. The main
features of the Kondo effect in quantum dots are a zero-bias conductance 
resonance, its specific temperature dependence and a splitting in a
magnetic field. 

The novel feature of realizing  Kondo physics in a 
semiconductor quantum dot is the possibility to apply a finite
bias $V_{sd}$ across the sample, which is not possible in the case of an ordinary
metal. In contrast to previous work we focus on the nonequilibrium properties
of a quantum dot in the Kondo regime with the tunneling barriers connected
to the leads with different strengths. 

The Kondo effect leads to an enhanced local density of states 
of the quantum dot at the Fermi levels of the electron reservoirs.
In the case of symmetric barriers this results in an enhanced conductance 
at zero bias which is rapidly decreasing for $V_{sd} \neq 0$.
Previous measurements~\cite{davidgg,crone} have been carried out
with symmetric barriers, i. e. they met the condition 
$\Gamma_L = \Gamma_R$, where $\Gamma_{L,R}$ denote the 
tunnel couplings of the dot to the left and right barrier, respectively.
However, the very essence of the Kondo resonance lies in the fact that 
it is pinned to the Fermi level of each contact which might also lead
to the occurence of a conductance anomaly at non-zero bias.
In related work on single charge traps in tunnel junctions a bias
splitting was found~\cite{ralph}, while it was not perfectly clear how
the barrier transmission of the electron trap was involved.

Here, we demonstrate that by tuning a quantum dot into strongly asymmetric coupling 
an offset of the Kondo conductance resonance to non-zero bias
can be observed at $B = 0~T$. This effect is due to the pinning  
of the resonance to the Fermi level of the more strongly coupled lead. 

Following the idea of Kondo physics in a dot we build a small quantum
dot similar to the one fabricated by Goldhaber-Gordon~{\it et al.}~\cite{davidgg}.
Metallic gates are deposited on the top of an AlGaAs/GaAs heterostructure
with its two-dimensional electron gas (2DEG) 50~nm below the surface.
The 2DEG has a low temperature mobility of $8\times 10^5$~cm$^2/$Vs 
and a density of $1.6\times 10^{11}$~cm$^{-2}$. 
A small electron island is formed within the 2DEG by
applying negative voltages $V_g$ to the gates (cf. lower inset of Fig.~1(a)).
The measurements are performed in a 
dilution refrigerator at 25~mK with standard lock-in techniques. 

As is shown in Fig.~1(a) we first concentrated on the conventional Coulomb blockade (CB)
with symmetric barriers. The level diagram for this situation is
sketched in the upper inset: At non-zero bias electrons can tunnel through the
ground state $\epsilon_0$ or excited states of the dot if these are attainable in the
energy window opened between the chemical potentials $\mu_{L,R}$ of the leads.
The CB diamond displayed in the left of
Fig.~2 allows the determination of the total capacitance of the dot~\cite{weis}. 
Comparing to the capacitance of a metallic disc, we obtain $r \approx 70$~nm
as the radius of the island corresponding to an electron number of
20 and a charging energy of $U = 2.7$~meV. 
The spin-degenerate mean level spacing can be estimated to be $\Delta = \frac{2 \hbar^2}{m^* r^2}
\approx 500~\mu$eV~\cite{revdots2}.
From the temperature dependence of the lineshape the total intrinsic
width of the resonances  is found to be $\Gamma=\Gamma_L+\Gamma_R=100~\mu$eV. 
We determine the electron temperature
to be $T_e \approx 100-120$~mK since the linewidth
saturates at this temperature. 

In order to tune into the asymmetric Kondo regime we adjusted the 
transparencies of the barriers 
by a plunger gate close to the right tunneling contact (gate \#2 in
the inset of Fig.~1(a)). As required for the Kondo effect 
the tunneling barriers are opened  
first by reducing the applied voltage to the gates \#1. Asymmetry is then achieved
by adjusting the right barrier opacity with the additional gate \#2.  
Having prepared the dot in such a way,
we obtain the peak structure shown in Fig.~1(b). The peaks at 
$V_{sd}=|\mu_L-\mu_R|=0$ form pairs for each state with an odd number of electrons
and the conductance background in the valley between the peaks is increased.
This peak structure is due to the spin-degeneracy of the energy levels
of the quantum dot. The energy required for tunneling onto the quantum dot
with an odd number of electrons is determined by the charging energy $U$.
In contrast, an electron tunneling onto the dot with an even electron number 
additionally has to pay the level spacing $\Delta$ as it has to occupy a different 
spatial quantum state. This uniquely identifies the regions of odd electron number
in which Kondo phenomena are expected to be observable.

In our case, the barriers define
different tunneling rates for the right and left lead. The inset of
Fig.~1(b) depicts this relation: 
Under strong coupling conditions in addition to the ground state $\epsilon_0$
a Kondo resonance appears in the density of states (DOS) of the quantum dot.
At non-zero bias this Kondo resonance is split into two resonances each of which is
pinned to the chemical potential of the respective lead. Due to the different 
couplings of the quantum dot to the leads one of the resonances
is enhanced whereas the other is suppressed. To obtain this enhancement, 
additionally a strong energy dependence of the barrier transparencies
has to be assumed, i.e. $\Gamma=\Gamma(E)$. 
The asymmetry and energy dependence of the tunneling 
strengths finally results in an enhanced conductance in the CB region 
at non-zero bias. 
This is also supported by first numerical simulations~\cite{schoeller99}. 

The width of the conductance resonances as determined from their 
temperature dependence now is $\Gamma \approx 200~\mu$eV -- at the same time
the peak height is reduced. As the peak height 
$\propto \Gamma_L \Gamma_R/(\Gamma_L + \Gamma_R)$ is dominated
by the more opaque barrier, this clearly indicates that the tunnel barriers
are now tuned into an asymmetric situation.
This is verified by estimations of $\Gamma_{L,R}$ from nonlinear transport measurements:
The tunnel couplings to the leads can be obtained via $\Delta I \approx e/h \Gamma_{L,R}$,
where $\Delta I$ is the increase in current when an additional transport channel
is opened in resonance with the source or drain contact, respectively~\cite{weis}.
This yields $\Gamma_L \approx 170~\mu$eV and $\Gamma_R \approx 80 ~\mu$eV which is
consistent with the temperature dependent measurements:
the asymmetry of the tunnel barriers is therefore $\Gamma_L/\Gamma_R \approx 2:1$ and 
the charging energy changes to $U \approx 1.5$~meV due to the previous retuning of the gates.
Accordingly, the Kondo temperature
$k T_K \approx (U \Gamma)^{1/2} \exp (-\pi |\mu_{L,R}-\mu_{dot}|/2 \Gamma)$~\cite{theo2} 
varies from a minimum value of 20~mK in the valley
to 6~K close to the conductance peaks \cite{chemdot}. 
$kT_K$ is equal to the width of the Kondo resonance. 
Comparing these values to $T_e$ we find $\Gamma/k T_e \approx
20$ and $T_K/T_e \approx 0.2 - 50$. From this we can deduce -- with regard to the calculations
in Ref. \cite{theo2} -- that we are well in the Kondo regime.

The comparison between the conventional CB and the Kondo
regime in non-equilibrium is given in the grayscale-plot of the conductance  
in Fig.~2: In (a) the well-known CB diamond structure is clearly seen. The
excited states emerge in the single electron tunneling regions, also revealing lines
of negative differential conductance (NDC)~\cite{weis}. It is commonly
assumed that these lines are caused by spin blockade~\cite{haeusler} being
a consequence of the spin selection rules for transport spectroscopy in a quantum dot.
In (b) the coupling of the dot to the leads 
is strongly enhanced and asymmetric, resulting
in a strong deviation from the diamond structure and the appearance of a 
conductance resonance at small negative bias. This resonance is well pronounced for
the upper pair of peaks, while it is only weak for the pair of
peaks in the lower part of the plot. 

In Fig.~3(a) a close-up of the grayscale plot of Fig.~2(b) at the position 
of the resonance around $V_g = -460~$mV is shown: 
In the CB region around zero bias and between the conductance
peaks first-order tunneling processes are forbidden.
However, due to the Kondo resonance at the Fermi level of the
left barrier the conductance is enhanced on the negative bias side. 
The diamond shape is strongly distorted in the region where the quantum dot 
is occupied with an odd number of electrons whereas no effect is visible 
in the regions above and below where the dot contains an even 
number of electrons. In the latter regions spin flip processes
are inhibited and therefore no Kondo resonance emerges at the Fermi level.
Applying a magnetic field perpendicular to the plane of the sample of
$B = 0.5$~T only, the Kondo effect is quenched and the regular 
diamond pattern of CB reappears. A slight increase 
in magnetic field reduces the tunnel couplings such that the system is tuned 
out of the Kondo regime.

In Fig.~3(b) the evolution of the Kondo resonance with gate voltage 
is shown. From the slope $C_{res}/C_{gate} \approx 3.6$
of the linear shift of the resonance with gate voltage the capacitive 
coupling to the left reservoir can be estimated
to be $60$~aF. With an on peak resistance of $\approx 450$~k$\Omega$ this
yields an $RC$-time of $27$~ps corresponding to a decay width of
$150~\mu$eV which is fully consistent with our estimation of the
coupling to the left reservoir. In the lowest trace a parabola superposed 
by a Lorentzian has been used to fit the width of the resonance. We obtain 
$0.42$~meV, or $T_K=4.9$~K, which is in agreement with our earlier 
estimation of $T_K$. 
 
Fig.~4(a) shows the temperature dependence of a conductance trace taken at 
negative bias as marked in Fig.~3(b). As expected
in the Kondo regime, with lower temperatures the
peak positions slightly shift towards each other~\cite{crone}. At the same time,
the conductance in the valley between the peaks passes through a minimum
to increase again at the lowest temperature. The temperature
dependence of the valley conductance at the positions noted
in Fig.~3(b) is shown in Fig.~4(a): Whereas at negative bias Kondo-like
behavior is found, at zero bias an intermediate and at positive bias
a convential temperature dependence of the valley conductance is observed.

Consequently, we obtain information on the spin orientation of the ground state
in the CB regime as it is marked in Fig.~2(a). 
The positions of  the NDC regions are located at bias voltages larger than found 
usually. According to Ref.~\cite{haeusler} this indicates that the quantum dot 
has a rather one-dimensional character.
For such systems it is guaranteed that the ground-state has   
either zero total spin or spin $S = 1/2$ \cite{haeusler,haeusler2}.
The spin positions can be related to the positions in the 
Kondo regime as shown in Fig.~2(b). From this, we further conclude 
that the Kondo effect is mediated by  $S=1/2$-states and not by 
higher spin states, as it might generally be possible. 

In conclusion, we demonstrated the occurrence of the Kondo effect in a 
few-electron quantum dot in the case of asymmetric barriers. 
We observe a pinning of the Kondo resonance to the chemical potential 
of the left lead. This results in an offset of the conductance anomaly in the CB
region to non-zero bias. The position of the anomaly shifts capacitively
coupled to one of the reservoirs. The temperature dependence of the
valley conductance changes with the applied bias from Kondo-like to 
conventional behavior.
Finally, we were able to show that the Kondo ground state is 
given by a spin $S = 1/2$-state, instead of a more complex multiplet structure.

We like to thank D.~Goldhaber-Gordon, D.~Abusch-Magder, Ch.~Bruder,
W.~H\"ausler, H. Schoeller, J. K\"onig, J. von Delft,
D.~A.~Wharam and D.~Pfannkuche for helpful discussions
and R.~J.~Warburton for critical reading of the manuscript. This work was funded 
in part by the Deutsche Forschungsgemeinschaft within project SFB~348. \\


\figure{Fig.~1: 
(a) Quantum dot conductance resonances in the Coulomb blockade regime.
The conductance is displayed as a function of the gate voltage $V_g$ applied
to the gates \#1 (cf. lower inset). 
The upper inset schematically shows the effective density of states (DOS)
of the quantum dot at almost zero bias. 
The lower inset shows a schematic of the device. With the gates \#1 the 
quantum dot is defined,
whereas gate \#2 is used to change the transparency of the right tunneling barrier.
(b) Conductance resonances in the case of strong coupling to one lead:
due to the spin-degeneracy the resonances group into 
pairs. In the valleys between peaks, separated by the charging energy $U$, 
the quantum dot is occupied by an odd number of electrons. 
In the inset of (b) the effective DOS of the quantum dot with an odd number 
of electrons is sketched in a non-equilibrium Kondo situation with asymmetric
barriers. The Kondo resonance is pinned on the thin barrier side.
\label{one}

\figure{Fig.~2: 
Grayscale plot of the bias and gate voltage dependence of the conductance (log-scale) 
(a) in the conventional CB and (b) in the 
Kondo regime (black: $0~\mu$S $\leq dI/dV \leq 0.05~\mu$S, white: $dI/dV \geq 20~\mu$S or 
$dI/dV < 0~\mu$S). 
In (a) the diamond structure and some excited states are well pronounced. The white arrows
indicate regions of negative differential conductance related to spin blockade. 
The arrows on the left give the spin orientation, as derived from the meaurements
of the Kondo resonance.
In (b) the pair structure of the peaks and the Kondo resonance in the valley can clearly 
be seen. The diamond shape is strongly distorted in this case. 
The spin orientation is also given (see details in the text).
} 
\label{two}

\figure{Fig.~3: 
(a) Magnification of the Kondo resonance for the upper pair of peaks
in Fig.~2(b) in a grayscale plot (linear scale, black: $0~\mu$S, white: $8~\mu$S). 
The resonance is offset to the negative bias region.
(b) The boxed region of Fig.~3(a) displayed with line plots. The lowest trace
is taken at $V_g=-0.4534$~V, the uppermost at $V_g=-0.4634$~V.
The resonance close to zero bias shifts linearly with the gate voltage. A fit to the
resonance in the lowest trace yields a Kondo temperature of $4.9$~K.
}
\label{three}

\figure{Fig.~4:
(a): Temperature dependence of the conductance trace taken at $V_{sd}=-0.19$~mV
(indicated by the arrow in Fig.~3(b),
$-\bullet-: 160$~mK, ------ : 400 ~mK, $- - - : 500$~mK,$-\circ-: 800$~mK).
(b): Temperature dependence of the minimum conductance in the
valley between the peaks for negative, zero and positive bias taken at the 
points marked in Fig.~3(b). 
}
\label{four}


\begin{thebibliography}{10}
\bibitem{chemdot} The chemical potential of the quantum dot is defined as
$\mu_{dot}=E(N+1)-E(N)$, where $E(N)$ is the ground state of the $N$-electron 
quantum dot.
\end{thebibliography}
\end{document}